\documentclass[%
reprint,
superscriptaddress,
 prl,
 amsmath,
 amssymb,
 aps,
 longbibliography
]{revtex4-2}

\usepackage{cancel}
\usepackage{graphicx}
\usepackage{dcolumn}
\usepackage{bm}
\usepackage[usenames,dvipsnames]{color}
\usepackage{xcolor}
\usepackage{soul}
\usepackage{enumitem}
\usepackage{xspace}
\usepackage{booktabs}
\usepackage{lipsum}
\usepackage{siunitx} 
\usepackage[ISO]{diffcoeff}
\usepackage{subcaption}
\usepackage{ragged2e}
\usepackage{mathrsfs}
\usepackage{hyperref}
\usepackage[capitalise]{cleveref}
\usepackage[normalem]{ulem}
\usepackage{amsmath}
\usepackage{esvect} 
\usepackage{titlesec}
\usepackage{titletoc}
\usepackage{chngcntr}
\usepackage{fontawesome5}
\usepackage{mleftright}
\usepackage[nodayofweek]{datetime}
\usepackage{comment}

\DeclareSIUnit\year{yr} 
\def\arraystretch{2}\tabcolsep=10pt

\DeclareCaptionJustification{justified}{\justifying}
\definecolor{firebrick}{HTML}{B22222}
\definecolor{green2}{RGB}{89, 122, 0}
\definecolor{green3}{RGB}{13, 66, 14}
\definecolor{green4}{RGB}{38, 74, 0}

\hypersetup{
    colorlinks=true,       
    linkcolor=firebrick,          
    citecolor=firebrick,        
    filecolor=firebrick,      
    urlcolor=firebrick          
}

\definecolor{orcid-green}{RGB} {166, 206, 57}
\newcommand{\MYhref}[3][blue]{\href{#2}{\color{#1}{#3}}}%

\titleclass{\mysection}{straight}[\section]
\titleformat{\mysection}[runin]
  {\itshape}{\thesection}{}{}[.---]
\titlespacing{\mysection}{1em}{1em}{0em}

\newcommand\funop[1]{\mathop{{}#1}}
\newcommand{\dd}{\mathop{}\!\mathrm{d}}
\newcommand{\mrm}[1]{\mathrm{#1}}
\newcommand{\usim}{\mathord{\sim}}
\newcommand{\bvec}[1]{\boldsymbol{#1}}

\DeclareSIUnit\clight{c}

\begin{document}

\title{\boldmath Magnetic Levitation as a New Probe of Non-Newtonian Gravity}

\author{Dorian W.~P.~Amaral\,\MYhref[orcid-green]{https://orcid.org/0000-0002-1414-932X}{\faOrcid}}
\email{damaral@ifae.es}
\affiliation{Department of Physics and Astronomy, Rice University, MS-315, Houston, TX, 77005, U.S.A.}

\author{Tim M. Fuchs\,\MYhref[orcid-green]{}{\faOrcid}}
\affiliation{School of Physics and Astronomy, University of Southampton,
SO17 1BJ, Southampton, UK}

\author{Hendrik Ulbricht\,\MYhref[orcid-green]{https://orcid.org/0000-0003-0356-0065}{\faOrcid}}
\affiliation{School of Physics and Astronomy, University of Southampton,
SO17 1BJ, Southampton, UK}

\author{Christopher D.~Tunnell\,\MYhref[orcid-green]{https://orcid.org/0000-0001-8158-7795}{\faOrcid}}
\affiliation{Department of Physics and Astronomy, Rice University, MS-315, Houston, TX, 77005, U.S.A.}

\begin{abstract}
\noindent
We present MORRIS (Magnetic Oscillatory Resonator for Rare-Interaction Studies) and propose the first tabletop search for non-Newtonian gravity due to a Yukawa-like fifth force using a magnetically levitated particle. Our experiment comprises a levitated sub-millimeter magnet in a superconducting trap that is driven by a time-periodic source. Featuring short-, medium-, and long-term stages, MORRIS will admit increasing sensitivities to the force coupling strength $\alpha$, optimally probing screening lengths of $\lambda \sim 1\,\mrm{mm}$. Our short-term setup provides a proof-of-principle study, with our medium- and long-term stages respectively constraining $\alpha \lesssim 10^{-4}$ and $\alpha \lesssim 10^{-5}$, leading over existing bounds.  Our projections are readily recastable to concrete models predicting the existence of fifth forces, and our statistical analysis is generally applicable to well-characterized sinusoidal driving forces. By leveraging ultralow dissipation and heavy test masses, MORRIS opens a new window onto tests of small-scale gravity and searches for physics beyond the Standard Model.
\end{abstract}
\maketitle

\mysection{Introduction}
\label{sec:intro}

Despite its success at astrophysical scales, the theory of general relativity remains irreconcilable with quantum mechanics, and no theory of quantum gravity has emerged~\cite{Penrose:2014nha,Greenberger:2010df}. Precision tests of gravity at millimeter scales and below offer a promising window onto its potential quantum nature. Observed deviations from the Newtonian prediction---the non-relativistic limit of general relativity---would signal a departure from classical gravity and hint at novel physics beyond the Standard Model~\cite{Adelberger:2003zx,Will:2005va,Hossenfelder:2010zj}. These could manifest as violations of the inverse-square law (ISL) and shifts in the expected gravitational interaction strength.

Many beyond Standard Model scenarios lead to modifications of gravity at $\mathcal{O}(1\,\mrm{mm})$ scales~\cite{Adelberger:2003zx,Adelberger:2009zz}. For instance, string theory introduces extra spatial dimensions at these scales, with superstring theory predicting the presence of light moduli fields, both of which result in deviations from Newtonian gravity~\cite{Arkani-Hamed:1998jmv,Dimopoulos:1996kp,Dimopoulos:2003mw}. Meanwhile, incorporating additional gauge symmetries into the Standard Model can result in new massive mediators that propagate Yukawa-like fifth forces~\cite{Fischbach:1992fa}.
Laboratory searches for millimeter-scale Yukawa forces have primarily employed torsion balances and pendula~\cite{Geraci:2010ft,Kapner:2006si,Yang:2012zzb}, with levitated setups utilizing optically suspended microspheres that have pushed sensitivities into the micron regime~\cite{Blakemore:2021zna,Venugopalan:2024kgu}. However, magnetic levitation strategies are yet to be exploited in the search for a generic fifth force~\cite{PhysRevApplied.8.034002,Janse:2024kcn}.

Magnetic levitation setups offer ultralow mechanical dissipation with high quality factors ($Q\gtrsim 10^6$) and the ability to suspend millimeter-scale magnets. This makes them excellent force and acceleration sensors~\cite{Vinante2020,Gonzalez-Ballestero:2021gnu,Latorre:2022vxo,Hofer2023,Schmidt:2024hdc}, with sensitivities reaching values as low as $10^{-16}\,\mrm{N\,Hz^{-1/2}}$, allowing them to detect purely gravitational interactions~\cite{Fuchs:2023ajk}.  Unlike optical or electrostatic traps, magnetic levitation relies on passive Meissner repulsion to levitate heavy test masses, introducing virtually no active-feedback noise and boosting the signal for forces that couple proportionally to mass. These features make this technology a powerful choice for probing small deviations from Newtonian gravity and the signals predicted by many beyond Standard Model theories.

Despite its promise, magnetic levitation technology is only now being employed in tests of fundamental physics~\cite{Vinante:2022hnf,Bose:2023gwh}. Levitated magnets have recently been shown to be a leading option in the search for ultralight~\cite{Higgins:2023gwq,Li:2023wcb,Kilian:2024fsg,Amaral:2024rbj}  and ultraheavy~\cite{Qin:2025jun} dark matter, as well as for high-frequency gravitational waves~\cite{Carney:2024zzk}. At the interface of quantum theory and gravity, proposals have also been put forward to search for macroscopic superposition states~\cite{Bose:2023gwh} and to test MOND-like modifications of gravity~\cite{Timberlake:2021oqf,Milgrom:1983ca}.
However, unlike MOND, a generic Yukawa‐type parameterization makes no assumption on the acceleration scale at which new physics enters, providing a continuous framework for uncovering deviations from Newtonian gravity across all scales.

In this Letter, we introduce MORRIS (Magnetic Oscillatory Resonator for Rare-Interaction Studies) and evaluate the sensitivity of the state-of-the-art and beyond in magnetic levitation to non-Newtonian gravity due to an additional Yukawa potential. Our proposal employs a sub-millimeter magnetic particle trapped via the Meissner effect and driven by a time-periodic source.  We develop three experimental stages (short\nobreakdash-, medium\nobreakdash-, and long\nobreakdash-term configurations) and use a likelihood-based approach to project our sensitivities to the coupling strength $\alpha$ over screening length $\lambda$. By comparing our projections to existing fifth force searches, we demonstrate that MORRIS will be able to uniquely test small-scale gravity and provide broad access to physics beyond the Standard Model.

\mysection{Fifth Forces}
\label{sec:theory}

A Yukawa potential can be generically parameterized by a dimensionless strength parameter $\alpha$ and a screening length $\lambda$. The strength parameter can be positive or negative, leading to either an attractive or a repulsive interaction. Including gravity, the relevant potential for our experiment is
\begin{equation}
    U(r) \equiv - \frac{G_\infty m_s m_p}{r} \left(1 + \alpha e^{-r / \lambda}\right)\,,
\label{eq:pot}
\end{equation}
where $m_s$ and $m_p$ are the masses of the source and levitated particle, respectively, which are placed a relative distance $r$ away from each other. The constant $G_\infty$ is Newton's gravitational constant.

\begin{figure}[!t]
    \centering    \includegraphics[width=\columnwidth]{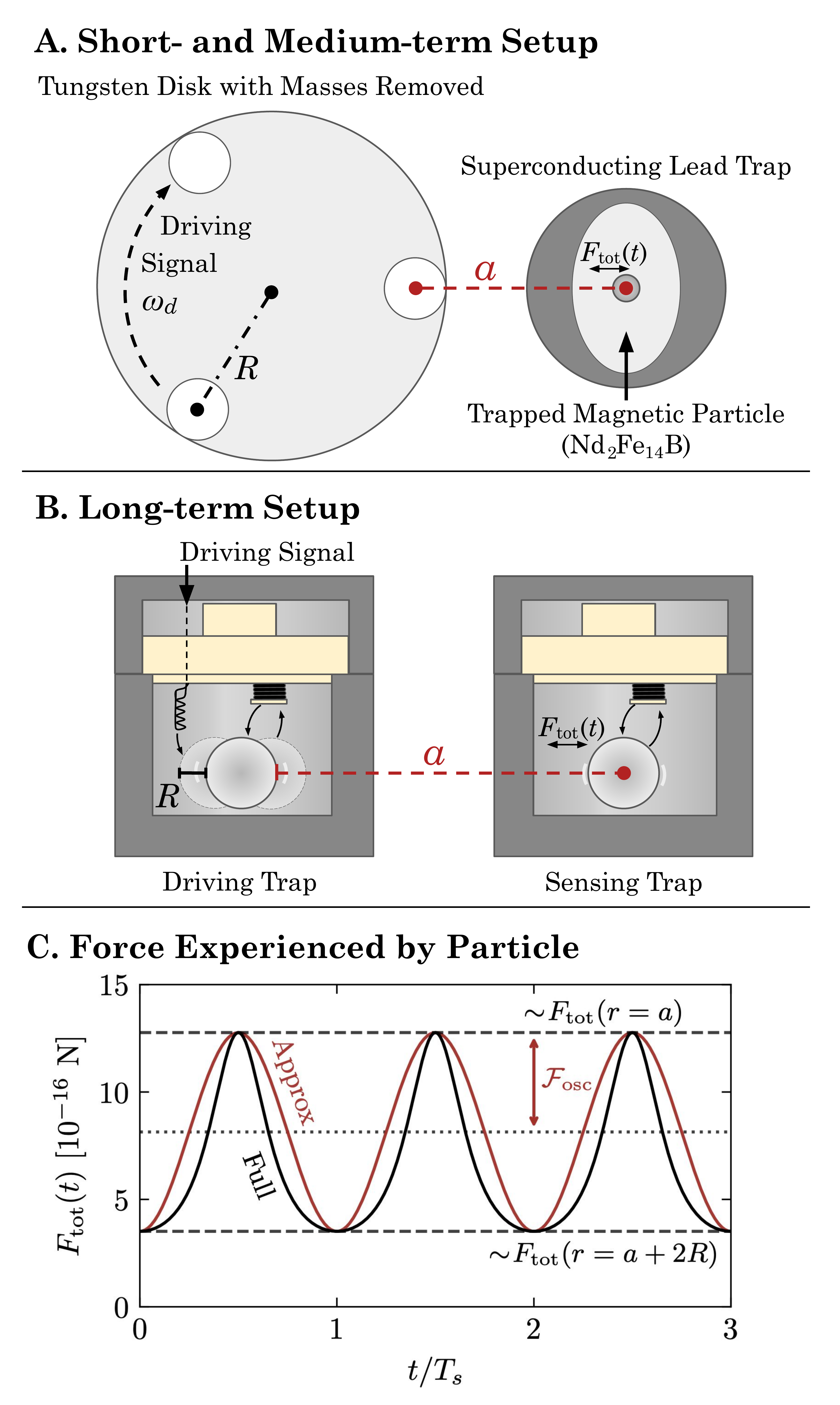}
    \caption{\textbf{A:}~Top view of the proposed short- and medium-term experimental setups for MORRIS, showing the mass-wheel and the trapped particle. We include the separation between the driving and test masses at closest approach $a$, the distance modulation $R$, the frequency of the drive $\omega_d$, and the total force experienced by the particle $F_\mrm{tot}(t)$. 
    \textbf{B:}~Side view of the proposed long-term setup, consisting of two traps. In the driving trap, an AC electromagnetic drive excites the trapped particle.
    \textbf{C:}~The force experienced by the sensing particle with signal period $T_s$. The dependence of this force to $a$ and $R$ is explained in Ref.~\cite{supp_mat}.}
    \label{fig:setup}
\end{figure}

From this potential, we may compute the total force using $\bvec{F}(\bvec{r}) = - \bvec{\nabla} U(r)$:
\begin{equation}
    \bvec{F}(\bvec{r}) = -\frac{G_\infty m_s m_p}{r^2} \left[1 + \alpha \left(1 + \frac{r}{\lambda}\right)e^{-r / \lambda}\right] \hat{\bvec{r}} \,.
    \label{eq:force}
\end{equation}
In our experiment, we generate this force via a time-periodic source. We achieve this by using either a spinning disk with three cylindrical masses removed in the case of our short- and medium-term setups, or a driven magnetically levitated particle within a second trap for our long-term stage. We illustrate these setups in \cref{fig:setup}. We place the source at a variable central displacement from the levitated particle $r_0$, representing the distance between the particle and the midpoint of the oscillation, possessing amplitude $R$. The most relevant physical scale is the closest approach distance between a source mass and the particle, $a(r_0) \equiv r_0 - R$. By varying this, we obtain sensitivity to ISL violations. 

At each central position $r_0$, we record the motion of the particle along a fixed sensitivity axis $\bvec{\zeta}$ using a superconducting pickup coil coupled to a SQUID. The time-dependent total force is then
\begin{equation}
    F_\mrm{tot}(t, r_0) \equiv \sum_{i=1}^{N_\mrm{mass}} \bvec{F}\textbf{(}\bvec{r}_i(t, r_0)\textbf{)} \cdot \hat{\bvec{\zeta}}\,,
\end{equation}
where the sum runs over the number of source masses $N_\mrm{mass}$. We align $\bvec{\zeta}$ with the $x$-axis of our experiment, along which we also place the oscillating source. 

To simplify our analysis, we decompose the total force into the sum of a constant component, which is irrelevant for our study, and an oscillatory component, $F_\mrm{osc}(t, r_0) \equiv \funop{\mathcal{F}_\mrm{osc}(r_0)} \sin(\omega_s t)$. The angular frequency $\omega_s = N_\mrm{mass} \omega_d$ is that of our periodic source, with $\omega_d$ the angular frequency of the drive. The amplitude $\mathcal{F}_\mrm{osc}(r_0)$ is given by half the difference between the maximal and minimal driving forces. We visualize the driving force in \cref{fig:setup} and validate our approximation in Ref.~\cite{supp_mat}. 

Ultimately, we conduct our analysis in Fourier space by analyzing the force power spectral density (PSD), $\mathcal{P}(\omega)$. Our signal manifests as a monochromatic peak at the angular frequency $\omega_s$, with a peak height proportional to $\mathcal{F}_\mrm{osc}^2(r_0)$~\cite{supp_mat}. The quantity of interest for our inferencing is the \textit{excess power} at this frequency normalized by the noise PSD $S_{FF}^\mrm{noise}(\omega_s)$. The normalized excess power at the central displacement $r_0$ after observing for a time $T_\mrm{obs}$ is given by the square of the quantity
\begin{equation}
    \kappa(r_0) \equiv \sqrt{\frac{\funop{\mathcal{F}_\mrm{osc}^2(r_0)} T_\mrm{obs}}{2 S_{FF}^\mrm{noise}(\omega_s)}}\,.
    \label{eq:kappa}
\end{equation}

\begin{figure}[!t]
    \centering    \includegraphics{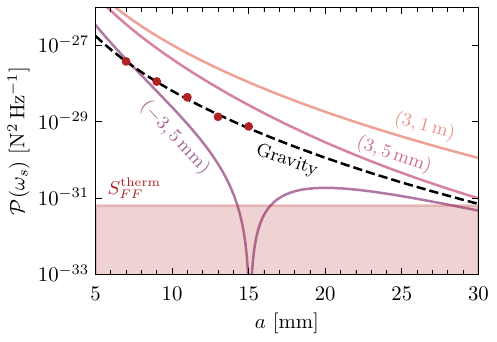}
    \caption{The force power spectral density of our levitated particle taken at the resonance frequency of the particle, $\mathcal{P}(\omega_s)$, with varying closest approach distance $a$. We assume our short-term configuration (c.f.~\cref{tab:configs}). Shown is the expected gravitational signal ($\alpha = 0)$ with five equally spaced data points (red markers). The fifth-force signals from three benchmark points are also shown, labeled according to $(\alpha, \lambda)$. The noise floor is assumed to be thermal, $S_{FF}^{\mrm{noise}}(\omega_s) = S_{FF}^\mrm{therm}$ (c.f.~\cref{eq:thermal_noise}).}
    \label{fig:sig_psd}
\end{figure}

We show the excess power, $\kappa^2(r_0) S_{FF}^\mrm{noise}(\omega_s)$, with varying closest approach distance in \cref{fig:sig_psd} for a choice of benchmark points $(\alpha, \lambda)$. We assume our short-term configuration, summarized in \cref{tab:configs}. We illustrate the gravitational signal ($\alpha = 0$), together with a set of simulated data points. We show the signals from a Yukawa force with parameters $(3, 1\,\mrm{m})$, $(3, 5\,\mrm{mm})$, and $(-3, 5\,\mrm{mm})$.  While these benchmark points are already firmly excluded, we visualize them to clearly illustrate the effect of a fifth force in our experiment. Moreover, though we display the signals over the full experimental range, we opt to measure them over a shorter scale to improve our overall signal-to-noise ratio. The noise is assumed to be thermal, $S_{FF}^{\mrm{noise}}(\omega_s) = S_{FF}^\mrm{therm}$, as given in \cref{eq:thermal_noise} and justified below.

For the first benchmark point, the fifth force is not effectively screened, resulting in a signal that does not observably violate the ISL; it instead appears as a scaling of the gravitational signal. For the remaining benchmark points, ISL violations are expected, with the latter featuring a repulsive force that can lead to a cancellation of the gravitational signal. This can be seen at $a \approx 15\,\mrm{mm}$. Since systematic uncertainties can scale our signal but are unlikely to mimic the spectral features induced by ISL violations, we are most sensitive to screening lengths of the order of our shortest experimental scale, $\lambda \sim a_\mrm{min} \sim\mathcal{O}(1\,\mrm{mm}\text{--}10\,\mrm{mm})$.

\mysection{Experiment}
\label{sec:experiment}
We present our proposed experiment, MORRIS, in \cref{fig:setup}. At its core is a composite particle consisting of a magnetic core and a non-magnetic weight that is levitated within a superconducting trap and driven by an external periodic force. We will use $\mrm{Nd_2Fe_{14}B}$ magnets due to their high remnant magnetization to aid in signal detection, silica to add mass without increasing eddy-current damping, and lead traps for their high type-I transition temperature (7.2 K). The magnetic trapping mechanism yields high quality factors $Q$ (in excess of $10^7$) at low resonance frequencies $f_0$ ($\usim 10\,\mrm{Hz}$)~\cite{PhysRevApplied.13.064027}.
Low-frequency operation enables the generation of a time-varying force gradient with a high modulation depth, leading to an increased signal. 
The gradient is sourced near the resonance frequency of the trapped particle, such that its motional amplitude steadily increases. 
This will be recorded via a pickup coil coupled to a SQUID---the conversion from SQUID signal to motion will be made either through thermal noise calibration or by injecting an external flux signal~\cite{Fuchs:2023ajk}.

Our experiment features short\nobreakdash-, medium\nobreakdash-, and long\nobreakdash-term stages that will allow us to search for non-Newtonian gravity with increasing sensitivities. They are expected to take one, two, and five years to construct, respectively. We summarize them in \cref{tab:configs}, providing technical details in Ref.~\cite{supp_mat}. 

\begin{table}[b]
\renewcommand{\arraystretch}{1.5}
    \centering
    \begin{tabular*}{\columnwidth}{@{\extracolsep{\fill}}lccc}
    \toprule\midrule
     \textbf{MORRIS} & Short & Medium & Long\\
    \midrule
     $m_p~\mrm{[mg]}$ & $4$ & $40$ & $400$ \\
     $m_s~\mrm{[mg]}$ & $210$ & 
     $350$ & $400$ \\
     $r_0^\mrm{min}~\mrm{[mm]}$ & 19.5 & 19.5 & 3\\
    $a_\mrm{min}~[\mrm{mm}]$ & 7 & 7 & 2\\
     $R~\mrm{[mm]}$ & 12.5 & 12.5 & 1\\
     $T_\mrm{obs}~\mrm{[h]}$ & $20$ & $1000$ & $1000$ \\
     $T~\mrm{[K]}$ & $4$ & $1$ & $0.1$ \\
     $f_0~[\mrm{Hz}]$ & 35 & 35 & 150\\
     $N_\mrm{mass}$ & 3 & 3 & 1\\
     $Q$ & $10^6$ & $10^7$ & $10^7$ \\
     $\sigma_\xi$ & $10\%$ & $1\%$ & $0.1\%$\\
     \midrule\bottomrule    
    \end{tabular*}
    \caption{The short-, medium-, and long-term configurations for our experiment, MORRIS. Shown is the the mass of the levitated particle ($m_p$), the mass of a single source mass ($m_s$), the minimum central displacement ($r_0^\mrm{min}$), the minimum closest approach distance ($a_\mrm{min}$), the motional amplitude ($R$), the observation time for a single data point ($T_\mrm{obs}$), the cooling temperature ($T$), the resonance frequency of the particle ($f_0$), the number of driving masses ($N_\mrm{mass}$), the quality factor ($Q$), and the uncertainty in the pull parameter ($\sigma_\xi$).}
    \label{tab:configs}
\end{table}

In the short- and medium-term setups, the force is sourced by a high-density solid disk of radius $1.25\,\mrm{cm}$ with three cylindrical masses removed. This density modulation sources a gradient when the disk is spun, such that the driving force varies with a frequency matching the resonance frequency of our particle. We will use tungsten for its high density, ease of machining, and commercial availability. Unlike in previous work, we will place the driving wheel inside of the cryogenic cooling system to minimize the shortest approach distance and attain sensitivities down to $\lambda \sim 1\,\mrm{mm}$~\cite{Fuchs:2023ajk}.

Our proposed long-term realization instead utilizes a direct drive of a particle suspended within a second trap. The second particle is driven either through an electrostatic interaction or via magnetic coils. This setup is similar to that proposed for precision measurements of the gravitational constant and in earlier work performed by the group~\cite{headley2025quantummetrologynewtonsconstant,Timberlake:2021oqf}. The geometry of this configuration allows us to achieve a smaller closest approach distance, halving $a_\mrm{min}$, while still guaranteeing adequate shielding from external electromagnetic forces. To minimize this distance, we will reduce the size of the traps from $\usim 1\,\mrm{cm}$ down to $2.5\,\mrm{mm}$ along the minor axis~\cite{Fuchs:2023ajk}.

During the experiment, sufficiently shielding the particle from external disturbances is crucial. This includes electromagnetic shielding and vibration isolation. Such shielding has been shown to adequately isolate a $0.4\,\mrm{mg}$ mass to a translational mode temperature of below $4\,\mrm{K}$~\cite{Fuchs:2023ajk}, placing the excess noise below the thermal noise for our short-term stage. Higher levels of vibration isolation have been reported by the LIGO collaboration and implementing similar levels of isolation would allow for thermally limited operation down to $100\,\mrm{mK}$~\cite{Matichard:2015eva}. Moreover, we will optimize the coupling to the particle's motion  such that the SQUID-readout noise is below the thermal noise floor, achieving this by increasing the number of pick-up coil windings and the geometrical coupling to the particles motion. Fiducializing to our short-term setup and using $\gamma \equiv \omega_0 / Q$ as our mechanical dissipation rate, our expected thermal noise is
\begin{align}
\label{eq:thermal_noise}
    &S_{FF}^\mrm{therm} = 4 m_p k_B T \gamma\\
    &\approx 10^{-31}\,\mrm{N^2\,Hz^{-1}}\left(\frac{m_p}{4\,\mrm{mg}}\right)\left(\frac{T}{4\,\mrm{K}}\right)\left(\frac{f_0}{35\,\mrm{Hz}}\right)\left(\frac{10^6}{Q}\right)\,.\nonumber
    \end{align}

Systematic uncertainties in our experiment can arise from how precisely we are able to measure masses, displacements, temperatures, and frequencies. In our statistical analysis, we account for them by introducing a pull parameter $\xi$ in our inferencing pipeline. This parameter serves to scale the signal-to-noise ratio and can be interpreted as its fractional uncertainty. A justification for the pull parameter uncertainties is given in Ref.~\cite{supp_mat}.

A key advantage of our experimental strategy is that we have excellent control over our source. Not only can we generate a driving force at a well-characterized frequency, such that we can target the signal in Fourier space, but we can also instrument the source masses to have precisely known values and vary the separation between the source and test particle. Placing the source in the vacuum  greatly reduces the minimum separation between the test and source masses, giving us sensitivity to smaller screening lengths $\lambda$.

Furthermore, measuring the force at varying separations gives us sensitivity to ISL violations. We will therefore make five equally spaced measurements, beginning from the minimum allowed central displacement of each setup $r_0^\mrm{min}$: $19.5\,\mrm{mm}$ for our short- and medium-term stages and $3\,\mrm{mm}$ for our long-term goal. We will displace the source $2\,\mrm{mm}$ at a time in the first two setups, improving this to $1\,\mrm{mm}$ in the long term. Remaining close to the minimal displacements gives us larger signal-to-noise ratios. Five data points are sufficient to give us competitive sensitivities when spectral information is available while maintaining a realistic total expedition time of $\usim 5T_\mrm{obs}$. 
From an experimental perspective, sampling at different separations allows for the rejection of systematic errors in both the closest separation $a$ and the force modulation depth by matching the curvature of the results.

\mysection{Analysis and Results}
\label{sec:anal}
 
Our expected signal is a monochromatic peak in the force PSD~\cite{supp_mat}. The likelihood function governing the noise-normalized excess power is a non-central $\chi^2$ function, $\chi^2_\mrm{nc}$, with $k=2$ degrees of freedom~\cite{1975ApJS...29..285G}. For the $i^\mrm{th}$ data point $p_i$, taken at the central displacement $r_{0,i}$, the non-centrality parameter is $\kappa^2_i \equiv \kappa^2(r_{0,i})$. To account for potential systematic uncertainties in both the signal and the noise, we include a nuisance pull parameter $\xi$ in our analysis that we ultimately profile out. This enters as a scaling of the non-centrality parameter, such that $\kappa_i \rightarrow (1 + \xi) \kappa_i$. We take $\xi \sim \mathcal{N}(0, \sigma_\xi^2)$, where $\sigma_\xi$ is its uncertainty. 

Since each data point is identically and independently distributed, the total likelihood is the product of the likelihood of each individual data point. Including the penalty term for the pull parameter, our likelihood given the observed data vector $\bvec{p} \equiv \{p_i\}$ is
\begin{equation}
\begin{split}
    \mathcal{L}(\alpha, \lambda, \xi; \bvec{p}) &\equiv \prod_{i=1}^{N_\mrm{data}} \funop{\chi^2_\mrm{nc}\left(p_i; (1 + \xi)^2\kappa_i^2 , k=2\right)} \\
    &\times \frac{1}{\sqrt{2\pi}\sigma_\xi} \exp\left(- \frac{\xi^2}{2\sigma_\xi^2}\right)\,,
\end{split}
\label{eq:lik}
\end{equation}
where $N_\mrm{data} = 5$ is the number of measurements. 
Using this likelihood, we define the profile likelihood ratio $\lambda_\mrm{PLR}(\alpha, \lambda=\lambda_0; \bvec{p})$, which profiles away the pull parameter. We then construct the two-sided test statistic $t_\alpha \equiv - 2 \ln \left[\funop{\lambda_\mrm{PLR}(\alpha, \lambda=\lambda_0; \bvec{p})}\right]$ and use it to draw our projected sensitivities~\cite{Cowan:2010js}. We detail our inference pipeline in Ref.~\cite{supp_mat}, which is generally applicable to searches for well-characterized sinusoidal driving forces.

To calculate our projections, we simulate $10^3$ pseudo-datasets according to a gravity-only signal. For each $\lambda = \lambda_0$, we scan over $\alpha$ to find our limit, which occurs when $t_\alpha \approx 3.84$ at the $95\%$ confidence level (CL)~\cite{supp_mat}. This procedure gives us a distribution of limits, the median of which we take as our best estimator of the limit. We also use these distributions to define $1\sigma$ limit bands. Since our projections for positive and negative $\alpha$ are similar, we only visualize our positive sensitivities, providing the full limit landscape in Ref.~\cite{supp_mat}.

\begin{figure}[t]
    \centering    \includegraphics{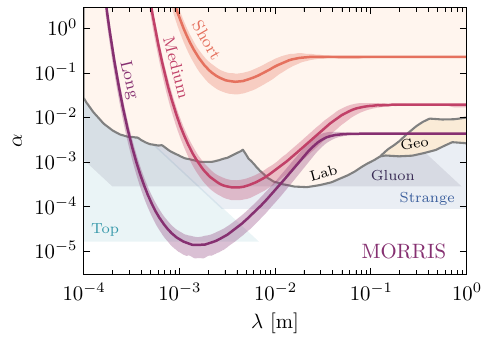}
    \caption{The $95\%$ confidence level projected limits on positive values of $\alpha$ with scale $\lambda$ from MORRIS. Shown are the limits from our short-, medium-, and long-term configurations, summarized in \cref{tab:configs}. The bands around each limit illustrate their $1\sigma$ range. The solid regions indicate the existing fifth-force limits from laboratory~\cite{Hoskins:1985tn,Kapner:2006si,Yang:2012zzb,Tan:2020vpf} and geophysical~\cite{Adelberger:2003zx} experiments. The transparent regions show the preferred regions for light moduli arising from supersymmetric string theory~\cite{Dimopoulos:1996kp,Dimopoulos:2003mw}.}
    \label{fig:limits}
\end{figure}

We show our projected $95\%$ CL limits in \cref{fig:limits}. All limits exhibit the same characteristic behaviors with varying screening length $\lambda$. We achieve optimal sensitivity when $\lambda \sim a_\mrm{min}$, the minimum closest approach distance of a source mass. This is because the amplitude $\mathcal{F}_\mrm{osc}$ is highly separation dependent, giving the data point at the closest separation the largest signal-to-noise ratio. This leads to a turnover at $\lambda \sim 7\,\mrm{mm}$ for the short- and medium-term realizations and at $\lambda \sim 2\,\mrm{mm}$ for the long-term stage (a similar scale to this was found in Ref.~\cite{Timberlake:2021oqf}). Since our projections are based on a generic Yukawa parametrization, they can be readily recast to concrete models predicting the existence of fifth forces, such as novel light scalar or vector mediators~\cite{Adelberger:2003zx}.

Below these characteristic distances, the fifth force is effectively screened over the scale of our experiment, leading to an exponential suppression of the signal. Conversely, above these distances, the signal is well described by a scaling of the null gravitational signal, allowing the pull parameter $\xi$ to account for the observed effect. At these high $\lambda$ values, the limiting $\alpha$ is dictated by the uncertainty in the pull parameter $\sigma_\xi$, such that $\alpha \sim \sigma_\xi$. Analytical estimates of our sensitivities, which corroborate with our numerical results, are given in Ref.~\cite{supp_mat}.

We compare our projections to existing laboratory and geophysical constraints in the $\lambda$ range of interest. The former consists of torsion balance~\cite{Hoskins:1985tn,Kapner:2006si} and pendulum~\cite{Yang:2012zzb,Tan:2020vpf} experiments, while the latter includes precision measurements of Earth's gravitational acceleration~\cite{Adelberger:2003zx}. Our short-term sensitivity, which is based on an immediately realizable configuration, provides a proof-of-concept for our experiment. Following a set of simple upgrades, a medium-term experiment will have leading sensitivity and constrain $\alpha\lesssim 10^{-4}$ at a screening length of $\lambda \approx 4\,\mrm{mm}$. Our long-term experiment is sensitive to new parameter space over the screening length range $0.3\,\mrm{mm}$ to $10\,\mrm{mm}$, constraining $\alpha \lesssim 10^{-5}$ at $\lambda \approx 2\,\mrm{mm}$ and surpassing laboratory limits by two orders of magnitude. 

The final two stages of our experiment will test regions of parameter space preferred by supersymmetric string theory~\cite{Dimopoulos:1996kp,Dimopoulos:2003mw}. Superstring theory introduces light moduli fields with masses governed by the supersymmetry breaking scale $\mathcal{F}_s$. These new scalar fields can mediate Yukawa-like fifth forces that couple to ordinary matter, as is the case for Yukawa and gauge moduli. The impact of these moduli in the fifth-force landscape is typically visualized in the sub-millimeter regime by assuming a cutoff scale of $\mathcal{F}_s \sim (10\,\mrm{TeV})^2$; however, a higher scale of $\mathcal{F}_s \sim (2000\,\mrm{TeV})^2$ is also well motivated~\cite{Giudice:1998bp,Dimopoulos:2003mw}. This extends the moduli regions to millimeter scales, where our experiment is highly sensitive. We show the preferred regions for the gluon, strange quark, and top quark moduli.

\mysection{Conclusions}
\label{sec:conc}

We have introduced MORRIS (Magnetic Oscillatory Resonator for Rare-Interaction Studies) and evaluated the sensitivity of magnetic levitation technology to non-Newtonian gravity. Our proposed tabletop experiment features a periodic source effecting a driving force on a levitated permanent magnet inside a superconducting trap. We considered a general Yukawa-like fifth force, described by a coupling strength $\alpha$ and screening length $\lambda$, that induces violations of the inverse-square law. We developed a likelihood framework generally applicable to searches for well-characterized sinusoidal driving forces, using it to compute the projected sensitivities of short-, medium-, and long-term realizations of our experiment. Our medium- and long-term stages will achieve leading sensitivities at millimeter scales in $\lambda$ ($\alpha \lesssim 10^{-4}$ and $\alpha \lesssim 10^{-5}$, respectively), with the latter boasting two orders of magnitude greater sensitivity than existing limits. These results can be readily recast to concrete models predicting the existence of fifth forces. Our work demonstrates the promise of magnetic levitation in gravity tests at the quantum frontier and in broader searches for physics beyond the Standard Model.

We thank Mustafa Amin, Andrew Long, Louis Hamaide, and Juehang Qin for their valuable comments on the manuscript. DA thanks Juehang Qin for helpful discussions on our statistical analysis, as well as Mustafa Amin and Andrew Long for insightful discussions throughout this work. DA
and CT were jointly funded by Rice University and NSF
award 2046549. TF and HU were supported by the EU Horizon Europe EIC Pathfinder project QuCoM (10032223) and as part of the EU QUANT-ERA project LEMAQUME, as well as by the UK funding agencies EPSRC (grants EP/V000624/1, EP/V035975/1, EP/W007444/1, EP/X009491/1) and STFC via the Southampton IAA project DMLOMS, by the Leverhulme Trust (RPG-2022-57), by ESA as parts of the Payload masters competition and by UKSA as part of the IBF project I3F.

\nocite{Baxter:2021pqo,Cowan:2010js,Timberlake:2021oqf,Fuchs:2023ajk,PhysRevLett.132.133602}

\bibliography{biblio}

@article{Adelberger:2003zx,
    author = "Adelberger, E. G. and Heckel, Blayne R. and Nelson, A. E.",
    title = "{Tests of the gravitational inverse square law}",
    eprint = "hep-ph/0307284",
    archivePrefix = "arXiv",
    doi = "10.1146/annurev.nucl.53.041002.110503",
    journal = "Ann. Rev. Nucl. Part. Sci.",
    volume = "53",
    pages = "77--121",
    year = "2003"
}

@article{Adelberger:2009zz,
    author = "Adelberger, E. G. and Gundlach, J. H. and Heckel, B. R. and Hoedl, S. and Schlamminger, S.",
    title = "{Torsion balance experiments: A low-energy frontier of particle physics}",
    doi = "10.1016/j.ppnp.2008.08.002",
    journal = "Prog. Part. Nucl. Phys.",
    volume = "62",
    pages = "102--134",
    year = "2009"
}

@article{Amaral:2024rbj,
    author = "Amaral, Dorian W. P. and Uitenbroek, Dennis G. and Oosterkamp, Tjerk H. and Tunnell, Christopher D.",
    title = "{First Search for Ultralight Dark Matter Using a Magnetically Levitated Particle}",
    eprint = "2409.03814",
    archivePrefix = "arXiv",
    primaryClass = "hep-ph",
    month = "9",
    year = "2024"
}

@article{Arkani-Hamed:1998jmv,
    author = "Arkani-Hamed, Nima and Dimopoulos, Savas and Dvali, G. R.",
    title = "{The Hierarchy problem and new dimensions at a millimeter}",
    eprint = "hep-ph/9803315",
    archivePrefix = "arXiv",
    reportNumber = "SLAC-PUB-7769, SU-ITP-98-13",
    doi = "10.1016/S0370-2693(98)00466-3",
    journal = "Phys. Lett. B",
    volume = "429",
    pages = "263--272",
    year = "1998"
}

@article{Baxter:2021pqo,
    author = "Baxter, D. and others",
    title = "{Recommended conventions for reporting results from direct dark matter searches}",
    eprint = "2105.00599",
    archivePrefix = "arXiv",
    primaryClass = "hep-ex",
    doi = "10.1140/epjc/s10052-021-09655-y",
    journal = "Eur. Phys. J. C",
    volume = "81",
    number = "10",
    pages = "907",
    year = "2021"
}

@article{Blakemore:2021zna,
    author = "Blakemore, Charles P. and Fieguth, Alexander and Kawasaki, Akio and Priel, Nadav and Martin, Denzal and Rider, Alexander D. and Wang, Qidong and Gratta, Giorgio",
    title = "{Search for non-Newtonian interactions at micrometer scale with a levitated test mass}",
    eprint = "2102.06848",
    archivePrefix = "arXiv",
    primaryClass = "hep-ex",
    doi = "10.1103/PhysRevD.104.L061101",
    journal = "Phys. Rev. D",
    volume = "104",
    number = "6",
    pages = "L061101",
    year = "2021"
}

@article{Bose:2023gwh,
    author = "Bose, Sougato and Fuentes, Ivette and Geraci, Andrew A. and Khan, Saba Mehsar and Qvarfort, Sofia and Rademacher, Markus and Rashid, Muddassar and Toro\v{s}, Marko and Ulbricht, Hendrik and Wanjura, Clara C.",
    title = "{Massive quantum systems as interfaces of quantum mechanics and gravity}",
    eprint = "2311.09218",
    archivePrefix = "arXiv",
    primaryClass = "quant-ph",
    doi = "10.1103/RevModPhys.97.015003",
    journal = "Rev. Mod. Phys.",
    volume = "97",
    number = "1",
    pages = "015003",
    year = "2025"
}

@article{Carney:2024zzk,
    author = "Carney, Daniel and Higgins, Gerard and Marocco, Giacomo and Wentzel, Michael",
    title = "{Superconducting Levitated Detector of Gravitational Waves}",
    eprint = "2408.01583",
    archivePrefix = "arXiv",
    primaryClass = "hep-ph",
    doi = "10.1103/PhysRevLett.134.181402",
    journal = "Phys. Rev. Lett.",
    volume = "134",
    number = "18",
    pages = "181402",
    year = "2025"
}

@article{Cowan:2010js,
    author = "Cowan, Glen and Cranmer, Kyle and Gross, Eilam and Vitells, Ofer",
    title = "{Asymptotic formulae for likelihood-based tests of new physics}",
    eprint = "1007.1727",
    archivePrefix = "arXiv",
    primaryClass = "physics.data-an",
    doi = "10.1140/epjc/s10052-011-1554-0",
    journal = "Eur. Phys. J. C",
    volume = "71",
    pages = "1554",
    year = "2011",
    note = "[Erratum: Eur.Phys.J.C 73, 2501 (2013)]"
}

@article{Dimopoulos:1996kp,
    author = "Dimopoulos, S. and Giudice, G. F.",
    title = "{Macroscopic forces from supersymmetry}",
    eprint = "hep-ph/9602350",
    archivePrefix = "arXiv",
    reportNumber = "CERN-TH-96-47",
    doi = "10.1016/0370-2693(96)00390-5",
    journal = "Phys. Lett. B",
    volume = "379",
    pages = "105--114",
    year = "1996"
}

@article{Dimopoulos:2003mw,
    author = "Dimopoulos, Savas and Geraci, Andrew A.",
    title = "{Probing submicron forces by interferometry of Bose-Einstein condensed atoms}",
    eprint = "hep-ph/0306168",
    archivePrefix = "arXiv",
    reportNumber = "STANFORD-ITP-03-17",
    doi = "10.1103/PhysRevD.68.124021",
    journal = "Phys. Rev. D",
    volume = "68",
    pages = "124021",
    year = "2003"
}

@article{Fischbach:1992fa,
    author = "Fischbach, E. and Talmadge, C.",
    title = "{Six years of the fifth force}",
    doi = "10.1038/356207a0",
    journal = "Nature",
    volume = "356",
    pages = "207--214",
    year = "1992"
}

@article{Fuchs:2023ajk,
    author = "Fuchs, Tim M. and Uitenbroek, Dennis G. and Plugge, Jaimy and van Halteren, Noud and van Soest, Jean-Paul and Vinante, Andrea and Ulbricht, Hendrik and Oosterkamp, Tjerk H.",
    title = "{Measuring gravity with milligram levitated masses}",
    eprint = "2303.03545",
    archivePrefix = "arXiv",
    primaryClass = "quant-ph",
    doi = "10.1126/sciadv.adk2949",
    journal = "Sci. Adv.",
    volume = "10",
    number = "8",
    pages = "eadk2949",
    year = "2024"
}

@article{Geraci:2010ft,
    author = "Geraci, Andrew A. and Papp, Scott B. and Kitching, John",
    title = "{Short-range force detection using optically-cooled levitated microspheres}",
    eprint = "1006.0261",
    archivePrefix = "arXiv",
    primaryClass = "hep-ph",
    doi = "10.1103/PhysRevLett.105.101101",
    journal = "Phys. Rev. Lett.",
    volume = "105",
    pages = "101101",
    year = "2010"
}

@article{Giudice:1998bp,
    author = "Giudice, G. F. and Rattazzi, R.",
    title = "{Theories with gauge mediated supersymmetry breaking}",
    eprint = "hep-ph/9801271",
    archivePrefix = "arXiv",
    reportNumber = "CERN-TH-97-380",
    doi = "10.1016/S0370-1573(99)00042-3",
    journal = "Phys. Rept.",
    volume = "322",
    pages = "419--499",
    year = "1999"
}

@article{Gonzalez-Ballestero:2021gnu,
    author = "Gonzalez-Ballestero, C. and Aspelmeyer, M. and Novotny, L. and Quidant, R. and Romero-Isart, O.",
    title = "{Levitodynamics: Levitation and control of microscopic objects in vacuum}",
    eprint = "2111.05215",
    archivePrefix = "arXiv",
    primaryClass = "quant-ph",
    doi = "10.1126/science.abg3027",
    month = "11",
    year = "2021"
}

@inproceedings{Greenberger:2010df,
    author = "Greenberger, Daniel M.",
    title = "{The Disconnect Between Quantum Mechanics and Gravity}",
    booktitle = "{Quantum Physics And The Nature Of Reality: A conference in honour of John Polkinghorne's 80th birthday}",
    eprint = "1011.3719",
    archivePrefix = "arXiv",
    primaryClass = "quant-ph",
    month = "11",
    year = "2010"
}

@article{Higgins:2023gwq,
    author = "Higgins, Gerard and Kalia, Saarik and Liu, Zhen",
    title = "{Maglev for dark matter: Dark-photon and axion dark matter sensing with levitated superconductors}",
    eprint = "2310.18398",
    archivePrefix = "arXiv",
    primaryClass = "hep-ph",
    reportNumber = "FERMILAB-PUB-23-624-SQMS-V",
    doi = "10.1103/PhysRevD.109.055024",
    journal = "Phys. Rev. D",
    volume = "109",
    number = "5",
    pages = "055024",
    year = "2024"
}

@article{Hofer2023,
  title = {High-$Q$ Magnetic Levitation and Control of Superconducting Microspheres at Millikelvin Temperatures},
  author = {Hofer, J. and Gross, R. and Higgins, G. and Huebl, H. and Kieler, O. F. and Kleiner, R. and Koelle, D. and Schmidt, P. and Slater, J. A. and Trupke, M. and Uhl, K. and Weimann, T. and Wieczorek, W. and Aspelmeyer, M.},
  journal = {Phys. Rev. Lett.},
  volume = {131},
  issue = {4},
  pages = {043603},
  numpages = {7},
  year = {2023},
  month = {Jul},
  publisher = {American Physical Society},
  doi = {10.1103/PhysRevLett.131.043603},
  url = {https://link.aps.org/doi/10.1103/PhysRevLett.131.043603}
}

@article{Hoskins:1985tn,
    author = "Hoskins, J. K. and Newman, R. D. and Spero, R. and Schultz, J.",
    title = "{Experimental tests of the gravitational inverse square law for mass separations from 2-cm to 105-cm}",
    doi = "10.1103/PhysRevD.32.3084",
    journal = "Phys. Rev. D",
    volume = "32",
    pages = "3084--3095",
    year = "1985"
}

@inbook{Hossenfelder:2010zj,
    author = "Hossenfelder, Sabine",
    title = "{Experimental Search for Quantum Gravity}",
    booktitle = "{Classical and quantum gravity}: {Theory, Analysis and Applications}",
    eprint = "1010.3420",
    archivePrefix = "arXiv",
    primaryClass = "gr-qc",
    month = "10",
    year = "2010"
}

@article{Janse:2024kcn,
    author = "Janse, Martijn and Uitenbroek, Dennis G. and van Everdingen, Loek and Plugge, Jaimy and Hensen, Bas and Oosterkamp, Tjerk H.",
    title = "{Current experimental upper bounds on spacetime diffusion}",
    eprint = "2403.08912",
    archivePrefix = "arXiv",
    primaryClass = "quant-ph",
    doi = "10.1103/PhysRevResearch.6.033076",
    journal = "Phys. Rev. Res.",
    volume = "6",
    number = "3",
    pages = "033076",
    year = "2024"
}

@article{Kapner:2006si,
    author = "Kapner, D. J. and Cook, T. S. and Adelberger, E. G. and Gundlach, J. H. and Heckel, Blayne R. and Hoyle, C. D. and Swanson, H. E.",
    title = "{Tests of the gravitational inverse-square law below the dark-energy length scale}",
    eprint = "hep-ph/0611184",
    archivePrefix = "arXiv",
    doi = "10.1103/PhysRevLett.98.021101",
    journal = "Phys. Rev. Lett.",
    volume = "98",
    pages = "021101",
    year = "2007"
}

@article{Kilian:2024fsg,
    author = "Kilian, Eva and others",
    title = "{Dark Matter Searches with Levitated Sensors}",
    eprint = "2401.17990",
    archivePrefix = "arXiv",
    primaryClass = "quant-ph",
    doi = "10.1116/5.0200916",
    journal = "AVS Quantum Sci.",
    volume = "6",
    pages = "030503",
    year = "2024"
}

@article{Latorre:2022vxo,
    author = "Latorre, Mart\'\i{} Gutierrez and Higgins, Gerard and Paradkar, Achintya and Bauch, Thilo and Wieczorek, Witlef",
    title = "{Superconducting Microsphere Magnetically Levitated in an Anharmonic Potential with Integrated Magnetic Readout}",
    eprint = "2210.13451",
    archivePrefix = "arXiv",
    primaryClass = "quant-ph",
    doi = "10.1103/PhysRevApplied.19.054047",
    journal = "Phys. Rev. Applied",
    volume = "19",
    number = "5",
    pages = "054047",
    year = "2023"
}

@article{Li:2023wcb,
    author = "Li, Rui and Lin, Shaochun and Zhang, Liang and Duan, Changkui and Huang, Pu and Du, Jiangfeng",
    title = "{Search for Ultralight Dark Matter with a Frequency Adjustable Diamagnetic Levitated Sensor}",
    eprint = "2307.15758",
    archivePrefix = "arXiv",
    primaryClass = "astro-ph.CO",
    doi = "10.1088/0256-307X/40/6/069502",
    journal = "Chin. Phys. Lett.",
    volume = "40",
    number = "6",
    pages = "069502",
    year = "2023"
}

@article{Matichard:2015eva,
    author = "Matichard, F. and others",
    title = "{Seismic isolation of Advanced LIGO: Review of strategy, instrumentation and performance}",
    eprint = "1502.06300",
    archivePrefix = "arXiv",
    primaryClass = "physics.ins-det",
    doi = "10.1088/0264-9381/32/18/185003",
    journal = "Class. Quant. Grav.",
    volume = "32",
    number = "18",
    pages = "185003",
    year = "2015"
}

@article{Milgrom:1983ca,
    author = "Milgrom, M.",
    title = "{A Modification of the Newtonian dynamics as a possible alternative to the hidden mass hypothesis}",
    doi = "10.1086/161130",
    journal = "Astrophys. J.",
    volume = "270",
    pages = "365--370",
    year = "1983"
}

@article{Penrose:2014nha,
    author = "Penrose, Roger",
    editor = "Scardigli, Fabio and Nespoli, Matteo",
    title = "{On the Gravitization of Quantum Mechanics 1: Quantum State Reduction}",
    doi = "10.1007/s10701-013-9770-0",
    journal = "Found. Phys.",
    volume = "44",
    pages = "557--575",
    year = "2014"
}

@article{Qin:2025jun,
    author = "Qin, Juehang and others",
    title = "{Mechanical Sensors for Ultraheavy Dark Matter Searches via Long-range Forces}",
    eprint = "2503.11645",
    archivePrefix = "arXiv",
    primaryClass = "hep-ph",
    month = "3",
    year = "2025"
}

@article{Schmidt:2024hdc,
    author = "Schmidt, Philip and others",
    title = "{Remote sensing of a levitated superconductor with a flux-tunable microwave cavity}",
    eprint = "2401.08854",
    archivePrefix = "arXiv",
    primaryClass = "quant-ph",
    doi = "10.1103/PhysRevApplied.22.014078",
    journal = "Phys. Rev. Applied",
    volume = "22",
    number = "1",
    pages = "014078",
    year = "2024"
}

@misc{supp_mat,
  note = "See Supplemental Material at [URL will be inserted by publisher] for further details on our fifth force signal approximation, employed statistics, full limit landscape, analytical sensitivity estimates, and technical details of our experimental choices."
}

@article{Tan:2020vpf,
    author = "Tan, Wen-Hai and others",
    title = "{Improvement for Testing the Gravitational Inverse-Square Law at the Submillimeter Range}",
    doi = "10.1103/PhysRevLett.124.051301",
    journal = "Phys. Rev. Lett.",
    volume = "124",
    number = "5",
    pages = "051301",
    year = "2020"
}

@article{Timberlake:2021oqf,
    author = "Timberlake, Chris and Vinante, Andrea and Shankar, Francesco and Lapi, Andrea and Ulbricht, Hendrik",
    title = "{Probing modified gravity with magnetically levitated resonators}",
    eprint = "2110.02263",
    archivePrefix = "arXiv",
    primaryClass = "gr-qc",
    doi = "10.1103/PhysRevD.104.L101101",
    journal = "Phys. Rev. D",
    volume = "104",
    number = "10",
    pages = "L101101",
    year = "2021"
}

@article{Venugopalan:2024kgu,
    author = "Venugopalan, Gautam and others",
    title = "{Search for new interactions at the micron scale with a vector force sensor}",
    eprint = "2412.13167",
    archivePrefix = "arXiv",
    primaryClass = "hep-ex",
    month = "12",
    year = "2024"
}

@article{Vinante2020,
  title = {Ultralow Mechanical Damping with Meissner-Levitated Ferromagnetic Microparticles},
  author = {Vinante, A. and Falferi, P. and Gasbarri, G. and Setter, A. and Timberlake, C. and Ulbricht, H.},
  journal = {Phys. Rev. Appl.},
  volume = {13},
  issue = {6},
  pages = {064027},
  numpages = {13},
  year = {2020},
  month = {Jun},
  publisher = {American Physical Society},
  doi = {10.1103/PhysRevApplied.13.064027},
  url = {https://link.aps.org/doi/10.1103/PhysRevApplied.13.064027}
}

@article{Vinante:2022hnf,
    author = "Vinante, Andrea and Timberlake, Chris and Ulbricht, Hendrik",
    title = "{Levitated Micromagnets in Superconducting Traps: A New Platform for Tabletop Fundamental Physics Experiments}",
    doi = "10.3390/e24111642",
    journal = "Entropy",
    volume = "24",
    number = "11",
    pages = "1642",
    year = "2022"
}

@article{Yang:2012zzb,
    author = "Yang, Shan-Qing and Zhan, Bi-Fu and Wang, Qing-Lan and Shao, Cheng-Gang and Tu, Liang-Cheng and Tan, Wen-Hai and Luo, Jun",
    title = "{Test of the Gravitational Inverse Square Law at Millimeter Ranges}",
    doi = "10.1103/PhysRevLett.108.081101",
    journal = "Phys. Rev. Lett.",
    volume = "108",
    pages = "081101",
    year = "2012"
}

@ARTICLE{1975ApJS...29..285G,
       author = {{Groth}, E.~J.},
        title = "{Probability distributions related to power spectra.}",
      journal = {ApJS},
         year = 1975,
        month = jun,
       volume = {29},
        pages = {285-302},
          doi = {10.1086/190343},
       adsurl = {https://ui.adsabs.harvard.edu/abs/1975ApJS...29..285G}
}

@article{PhysRevApplied.8.034002,
  title = {Ultrasensitive Inertial and Force Sensors with Diamagnetically Levitated Magnets},
  author = {Prat-Camps, J. and Teo, C. and Rusconi, C. C. and Wieczorek, W. and Romero-Isart, O.},
  journal = {Phys. Rev. Appl.},
  volume = {8},
  issue = {3},
  pages = {034002},
  numpages = {10},
  year = {2017},
  month = {Sep},
  publisher = {American Physical Society},
  doi = {10.1103/PhysRevApplied.8.034002},
  url = {https://link.aps.org/doi/10.1103/PhysRevApplied.8.034002}
}

@article{PhysRevApplied.13.064027,
  title = {Ultralow Mechanical Damping with Meissner-Levitated Ferromagnetic Microparticles},
  author = {Vinante, A. and Falferi, P. and Gasbarri, G. and Setter, A. and Timberlake, C. and Ulbricht, H.},
  journal = {Phys. Rev. Appl.},
  volume = {13},
  issue = {6},
  pages = {064027},
  numpages = {13},
  year = {2020},
  month = {Jun},
  publisher = {American Physical Society},
  doi = {10.1103/PhysRevApplied.13.064027},
  url = {https://link.aps.org/doi/10.1103/PhysRevApplied.13.064027}
}

@misc{headley2025quantummetrologynewtonsconstant,
      title={Quantum Metrology of Newton's Constant with Levitated Mechanical Systems}, 
      author={Francis J. Headley and Alessio Belenchia and Mauro Paternostro and Daniel Braun},
      year={2025},
      eprint={2503.16215},
      archivePrefix={arXiv},
      primaryClass={quant-ph},
      url={https://arxiv.org/abs/2503.16215}, 
}

@article{PhysRevLett.132.133602,
  title = {Ultrahigh Quality Factor of a Levitated Nanomechanical Oscillator},
  author = {Dania, Lorenzo and Bykov, Dmitry S. and Goschin, Florian and Teller, Markus and Kassid, Abderrahmane and Northup, Tracy E.},
  journal = {Phys. Rev. Lett.},
  volume = {132},
  issue = {13},
  pages = {133602},
  numpages = {7},
  year = {2024},
  month = {Mar},
  publisher = {American Physical Society},
  doi = {10.1103/PhysRevLett.132.133602},
  url = {https://link.aps.org/doi/10.1103/PhysRevLett.132.133602}
}

@article{Will:2005va,
    author = "Will, Clifford M.",
    title = "{The Confrontation between general relativity and experiment}",
    eprint = "gr-qc/0510072",
    archivePrefix = "arXiv",
    doi = "10.12942/lrr-2006-3",
    journal = "Living Rev. Rel.",
    volume = "9",
    pages = "3",
    year = "2006"
}

\clearpage

\appendix

\onecolumngrid

\begin{center}
\large
\textbf{
    \textit{Supplemental Material} \\ Magnetic Levitation as a New Probe of Non-Newtonian Gravity
    }
    
\vspace{1.75ex}

Dorian W.~P.~Amaral, Tim M.~Fuchs, Hendrik Ulbricht, and Christopher D.~Tunnell

\vspace{2ex}

\end{center}

\twocolumngrid

\section{A.~Fifth Force Signal}
\label{sec:approx}

The total force imparted on our levitated particle due to the driving potential is given by the sum of the individual forces from each source mass. However, we pick out the signal at the fundamental frequency in Fourier space; this is given by the frequency at which the potential varies. This motivates us to decompose the full force into a constant component, $\mathcal{F}_0$, and an oscillatory component varying at the signal angular frequency $\omega_s$. We therefore approximate the total force at time $t$ and central displacement $r_0$ as
\begin{equation}
    F_\mrm{tot}(t, r_0) \simeq \mathcal{F}_0(r_0) + \mathcal{F}_\mrm{osc}(r_0) \sin(\omega_s t + \varphi)\,,
    \label{eq:force_approx}
\end{equation}
where $\varphi$ is a phase that ultimately does not enter our analysis, and so we set it to zero. The signal frequency is given by $\omega_s = N_\mrm{mass}\,\omega_d$, where $N_\mrm{mass}$ is the number of driving masses and $\omega_d$ is the driving frequency. We find that this approximation is appropriate when the amplitude of the oscillation $R$ is smaller than the relative central displacement of the source from the test mass, such that $R < r_0$. This is true in all of our cases.

We may write the constant and oscillatory components as functions of the force extrema. The former is given by the average of these extrema, whereas the oscillatory amplitude is equal to half their difference. Due to the symmetry of the problem, these extrema occur at time intervals of half the signal period, $T_s / 2$, with $T_s \equiv 2\pi / \omega_s$. We thus have that
\begin{equation}
\begin{split}
    \mathcal{F}_0(r_0) &=\frac{1}{2}\left[\funop{F_{\mrm{tot}}(T_s/2, r_0)} + \funop{F_{\mrm{tot}}(0, r_0)}\right]\,,\\
    \mathcal{F}_\mrm{osc}(r_0) &= \frac{1}{2}\left[\funop{F_{\mrm{tot}}(T_s/2, r_0)} - \funop{F_{\mrm{tot}}(0, r_0)}\right]\,.
\end{split}
\label{eq:force_scales}
\end{equation}
For a single source mass, the oscillation amplitude is simply the difference between the total force at the closest separation $a$ and farthest separation $a + 2R$.

In Fourier space, the relevant signal is the amplitude of the peak at the signal frequency. This is controlled by $\mathcal{F}_\mrm{osc}$ and the observation time $T_\mrm{obs}$; its value is given by
\begin{equation}
    P(r_0) = \frac{\funop{\mathcal{F}_\mrm{osc}^2(r_0)} T_\mrm{obs}}{2}\,.
    \label{eq:peak-height}
\end{equation}
However, in reality, the total force $F$ experienced by the levitated particle is a sum of the signal force supplied by our driving potential and random forces originating from environmental noise. At any given central displacement, we may write this as
\begin{equation}
    F(t) = F_\mrm{tot}(t) + F_\mrm{noise}(t)\,.
\end{equation}
We assume that the noise is white, admitting equal power across all frequencies. For white noise, the random force should be normally distributed as $F_\mrm{noise} \sim \mathcal{N}(0, \sigma^2)$, where $\sigma^2 \equiv \funop{S_{FF}^\mrm{noise}(\omega)} f_\mrm{samp}$ is the force noise variance in the time domain, $S_{FF}^\mrm{noise}(\omega)$ is the force noise power spectral density (PSD), and $f_\mrm{samp}$ is the sampling frequency. 

The total expected (one-sided) force PSD is given by the discrete Fourier transform
\begin{equation}
    \mathcal P(\omega) \equiv 2 \frac{(\Delta t)^2}{T_\mathrm{obs}} \Bigg| \sum_{n = 0}^{N - 1} \funop{F(t_n)} e^{i \omega n \Delta t} \Bigg |^2\,,
    \label{eq:periodogram-dft}
\end{equation}
where $\omega$ is the angular frequency, $N$ is the number of points sampled in the time domain, and $\Delta t \equiv T_\mathrm{obs} / N$ is the sampling time step. The factor of two accounts for the `folding' of the result from negative to positive frequencies to produce the one-sided periodogram. This allows us to define the noise-normalized periodogram at a particular frequency, $\mathcal{P}(\omega) / S_{FF}^\mrm{noise}(\omega)$. For a monochromatic force such as the one approximately arising from our source potential, its value at the signal frequency is controlled by the square of the dimensionless parameter
\begin{equation}
    \kappa(r_0) \equiv \sqrt{\frac{\funop{\mathcal{F}_\mrm{osc}^2(r_0)} T_\mathrm{obs}}{2 S_{FF}^\mrm{noise}(\omega_s)}}\,.
    \label{eq:kappa_app}
\end{equation}
This is the quantity governing our inferencing when searching for a fifth force.

\begin{figure}[t]
    \centering    \includegraphics[width=\columnwidth]{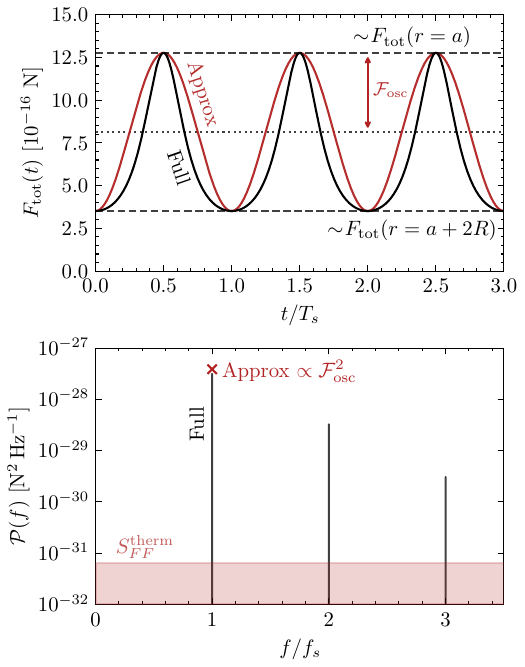}
    \caption{Comparisons between the gravitational signal ($\alpha=0$) from the full calculation and our approximation, \cref{eq:force_approx}. This is shown in the time domain (\textbf{top}) and in the power spectral density (\textbf{bottom}) using \cref{eq:periodogram-dft}. The oscillation amplitude $\mathcal{F}_\mrm{osc}$ is given by the difference between the total force at the closest separation $a$ and farthest separation $a + 2R$ for a single source mass. The approximation accurately captures the resonant peak, occurring at the signal frequency $f = f_s$ and which we use in our inferencing. The noise floor is assumed to be thermal and is computed in the main text. We use our short-term setup, taking $r_0 \approx 1.95\,\mrm{cm}$.}
    \label{fig:approx}
\end{figure}

We show the result of a simulation using the full force treatment and the approximation given by \cref{eq:force_approx} in \cref{fig:approx} for the gravity-only scenario ($\alpha = 0$). We visualize this using our short-term configuration in both the time domain and Fourier space; we use \cref{eq:periodogram-dft} for the latter. The PSD features the strongest peak at the fundamental frequency $f_s = \omega_s / (2\pi)$, with progressively weaker peaks at higher harmonics. Our approximation accurately captures the resonance peak, which is the part of the signal we plan to isolate in an experimental run. The value of the peak as predicted by our approximation is $\usim 1.23$ times larger than that from the full simulation for any choice of the fifth-force parameters $\alpha$ and $\lambda$. Since both the gravitational and new physics signals scale by the same factor, we rely on our approximation and do not expect this difference to make an appreciable impact on our results. We also show the thermal noise floor, as computed in the main text.

\section{B.~Inferencing Pipeline}

Using the likelihood given in the main text, $\mathcal{L}$, we define the profile-likelihood ratio
\begin{equation}
    \funop{\lambda_\mrm{PLR}(\alpha, \lambda=\lambda_0; \bvec{p})} \equiv \frac{\funop{\mathcal{L}\big(\alpha, \lambda_0, \hat{\xi}; \bvec{p}\,\big)}}{\funop{\mathcal{L}\big(\hat{\hat{\alpha}}, \lambda_0,  \hat{\hat{\xi}}; \bvec{p}\,\big)}}\,,
    \label{eq:plr}
\end{equation}
where $\bvec{p} \equiv \{p_i\}$ is our data vector. The single- and double-hats respectively denote the maximum likelihood estimators of the likelihood conditioned on a particular $(\alpha, \lambda)$ and of the unconditional likelihood.  We fix to one value of the screening length $\lambda_0$ and perform a raster scan of the parameter space over different values of $\lambda_0$~\cite{Baxter:2021pqo}.

From the profile-likelihood ratio, we construct the two-sided test statistic (TS)~\cite{Cowan:2010js}
\begin{equation}
 t_\alpha \equiv - 2 \ln \left[\funop{\lambda_\mrm{PLR}(\alpha, \lambda=\lambda_0); \bvec{p}}\right]\,.
 \label{eq:test-stat}
\end{equation}
We use the TS to assess whether the null hypothesis, defined according to conditional likelihood, is a good description of the observed data compared to the alternative hypothesis, defined as per the
maximized unconditional likelihood. We use a two-sided test statistic since $\alpha$ can take on both positive and negative values. A value of $0$ for $t_\alpha$ indicates perfect agreement with the null hypothesis,
whereas larger values indicate greater disagreement.   

To reject the null hypothesis, we define a threshold value for the TS above which the alternative hypothesis is preferred, $t_\alpha^\mrm{lim}$.
Specifically, if $f(t_\alpha|\alpha, \lambda_0)$ is the distribution of TS values that should arise from the null hypothesis, we define the $p$-value of the observed TS value  $t_\alpha^\mrm{obs}$ as
\begin{equation}
    p_\alpha \equiv \int_{t_\alpha^\mrm{obs}}^{\infty} f(t_\alpha | \alpha, \lambda_0) \dd t_\alpha\,.
    \label{eq:pval}
\end{equation}
This specifies the probability of finding a TS value at least as extreme as the one observed. To compute our $95\%$ confidence level (CL) limit, we then find that observed value for the TS such that $p_\alpha = 0.05$; this is $t_\alpha^\mrm{lim}$. 

To find $t_\alpha^\mrm{lim}$, we must know the distributions of the test statistic for each choice of null hypothesis, $f(t_\alpha|\alpha, \lambda_0)$. We generate these from a series of Monte Carlo (MC) simulations. For a large selection of null hypotheses, defined by the parameter space points $(\alpha, \lambda_0)$, we generate $10^6$ pseudo-datasets distributed according to our likelihood function at our chosen central displacements. For each dataset, we then compute the value of $t_\alpha$ from \cref{eq:test-stat}, generating the desired TS distributions. For $\alpha$ values close to the ones we can exclude, we find that the distributions of $t_\alpha$ are well described by the asymptotic result of Ref.~\cite{Cowan:2010js}, with $t_\alpha$  following a $\chi^2$ distribution with one degree of freedom. This gives us a limiting TS value of $t_\alpha^\mrm{lim} \approx 3.84$ for a $95\%$ CL limit.

While deviations from this asymptotic result occur, they do so for large absolute values of $\alpha$ that are already excluded. For these values, the distributions tend to favor lower values of the TS, shrinking the large $\alpha$ tail of the distribution. Ultimately, this leads to a lower value of $t_\alpha^\mrm{lim}$ for this region of parameter space, which we can nevertheless exclude with the higher value of $t_\alpha^\mrm{lim} \approx 3.84$. We thus take the asymptotic result for the TS distribution.

\section{C.~Full Limit Landscape}
\label{sec:app-limits}

\begin{figure}[t!]
    \centering
    \includegraphics{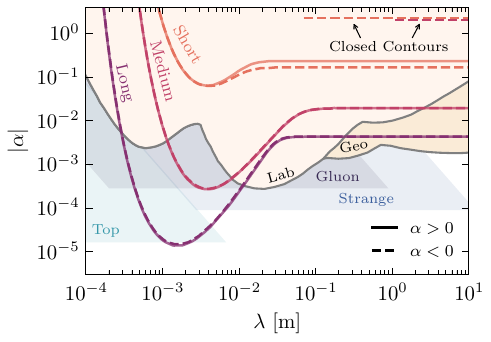}
    \caption{The $95\%$ confidence level projected limits on absolute values of $\alpha$ with screening length $\lambda$. We show the limits from our short-, medium-, and long-term configurations. The dashed lines above unity $|\alpha|$ are thin, closed contours where negative values of $\alpha$ are allowed. The existing constraints and theoretical regions are discussed in the main text.}
    \label{fig:lims-full}
\end{figure}

We provide our complete limit landscape for both positive and negative values of $\alpha$ in \cref{fig:lims-full}, in which we visualize their absolute values. The negative limits are almost identical for our medium- and long-term configurations. The short-term negative limit is slightly stronger than its positive counterpart.

For $\alpha \sim -1$, we find highly fine-tuned areas of parameter space where our limits form closed contours, indicating regions of allowed $\alpha$ values. In these regions, a repulsive fifth force exactly mimics the oscillatory gravitational signal but with a phaseshift of $\pi$ radians. In our Fourier space analysis, we only consider the amplitude of the signal, making us blind to phases. Thus, for these fine-tuned values, our signal appears to be consistent with the null result. However, since these regions are firmly within already excluded regions of parameter space, we do not study them further. 

\section{D.~Analytical Sensitivity Estimate}

We perform a sensitivity estimate by considering a single, oscillating source mass and by using an Asimov dataset that includes only the first data point~\cite{Cowan:2010js}.  For large non-centrality parameters $\kappa^2$, defined in \cref{eq:kappa_app}, we can approximate the non-central $\chi^2$ distribution to be Gaussian. Its mean $\mu$ and variance $\sigma^2$ are equal to those of the original distribution, such that $\mu \simeq \kappa^2$ and $\sigma^2\simeq 4\kappa^2$. We allow for the non-centrality parameter to be shifted by the pull parameter $\xi$, governing the effects of experimental systematics. Ignoring the penalty term on $\xi$, our approximate test statistic, formed from the log-likelihood ratio, is then
\begin{equation}
   t \simeq \frac{\left[\kappa_g^2 - (1+\xi)^2\kappa^2\right]^2}{4(1+\xi)^2\kappa^2}\,\,.
\end{equation}
We have taken our data point to be exactly consistent with a gravity-only signal, such that $p = \kappa_g^2$, with $\kappa_g$ given by setting $\alpha = 0$ in $\mathcal{F}_\mrm{osc}$. Setting the test statistic to $t_\mrm{lim} \approx 3.84$, as derived from \cref{eq:pval}, we can solve for the limiting non-centrality parameter. Since $\kappa^2 \gg t_\mrm{lim}$ for all of our cases, we find that
\begin{equation}
    \kappa^2_\mrm{lim} \simeq \frac{\kappa^2_g \pm 2 \sqrt{t_\mrm{lim}\kappa^2_g}}{(1 + \xi)^2}\,.
\end{equation}
We can then invert this to find a limit on $\alpha$.

To derive this limit, it is useful to define the quantities
\begin{equation}
\begin{split}
    \mathcal{D}(\lambda, r) &\equiv \frac{(1 + r/\lambda)\,e^{-r / \lambda}}{r^2}\,; \\  \Delta(\lambda) &\equiv \mathcal{D}(\lambda, a) - \mathcal{D}(\lambda, b)\,,
\end{split}
\end{equation}
where $a$ and $b$ are the closest and farthest separations of the test and source particles at a given central displacement $r_0$, respectively. We can then write the oscillatory force scale, defined in \cref{eq:force_scales}, as
\begin{equation}
    \mathcal{F}_\mrm{osc}(\alpha, \lambda) = \frac{1}{2} G_\infty m_s m_p \left[\funop{\Delta(\infty)} +\, \alpha \funop{\Delta(\lambda)}\right]\,.
\end{equation}
The gravitational force scale $\mathcal{F}_g$ is retrieved by setting $\alpha =0$. Using the definition of $\kappa$, we then find an approximate limit on $\alpha$,
\begin{equation}
    \alpha_\mrm{lim}(\lambda) \simeq\frac{1}{1 + \xi}\frac{\Delta(\infty)}{\Delta(\lambda)}\left(-\xi \pm \sqrt{t_\mrm{lim}\frac{2 S_{FF}^\mrm{noise}}{\mathcal{F}_g^2 T_\mrm{obs}}}\,\right)\,.
    \label{eq:alpha_approx}
\end{equation}

To compute this limit, we must choose a value for the pull parameter. However, the extent to which $\xi$ influences our statistics depends on the screening length $\lambda$. For $\lambda$ much larger than the minimum closest approach distance, $a_\mrm{min} \equiv r_{0}^\mrm{min} - R$, the fifth-force signal is a scaling of the gravitational one, such that systematics limit our overall sensitivity. Conversely, for smaller $\lambda$, the pull parameter plays a less significant role. Our true sensitivity thus lies between these two regimes.

\begin{figure}[t]
    \centering
    \includegraphics{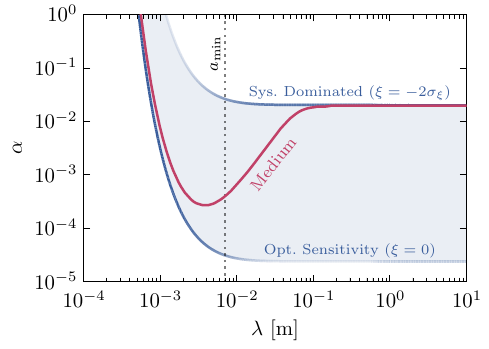}
    \caption{The $95\%$ confidence level projected limits on the strength parameter $\alpha$ with screening length $\lambda$ from our complete statistical treatment and analytical approximations. We visualize the result for our medium-term configuration. Below the minimum separation of $a_\mrm{min} \approx 7\,\mrm{mm}$, our approximation is based on the optimal sensitivity by setting the pull parameter $\xi$ to zero. Above this scale, systematics play a more dominant role, for which we set $\xi = -2 \sigma_\xi$. The true limit lies between these cases, highlighted by the shaded region. The transparency of the analytical estimates approximately indicates where each one should be trusted.}
    \label{fig:anal_alpha}
\end{figure}

These two cases are captured in the parentheses of \cref{eq:alpha_approx}. The first term controls the limit in the systematics-dominated regime, while the second term captures its behavior when systematics are not expected to play as important a role. In the former regime, the magnitude of the pull parameter should be $|\xi| \sim \sigma_\xi$. Setting $\xi$ to this, we find that the first term dominates since
\begin{equation}
    \sigma_\xi \gg \sqrt{t_\mrm{lim}\frac{2 S_{FF}^\mrm{noise}}{\mathcal{F}_g^2 T_\mrm{obs}}}\,.
\end{equation}
In the second case, we can set $\xi = 0$. Therefore, we may write our approximate limit as a piecewise function that reproduces these limiting regimes:
\begin{equation}
\hspace{0ex}
        \alpha_\mrm{lim}(\lambda) \simeq 
\begin{cases}
\pm 2 \sigma_\xi  / (1 \mp 2 \sigma_\xi)\quad &\text{if} \hspace{1.5ex} \lambda \gg a_\mrm{min} \\[1ex]
\pm \frac{\Delta(\infty)}{\Delta(\lambda)}\sqrt{t_\mrm{lim} \frac{2 S_{FF}^\mrm{noise}}{\mathcal{F}_g^2 T_\mrm{obs}}}  \quad &\text{if} \hspace{1.5ex} \lambda \lesssim a_\mrm{min}
\end{cases}\,.
\end{equation}
In the systematics-dominated regime, we set the pull parameter to $|\xi| = 2 \sigma_\xi$; this gives us good agreement with our numerical result. We compare our medium-term configuration result to our analytics in \cref{fig:anal_alpha}, where we find excellent agreement.

To aid with experimental design, we can expand the result of the non-systematics dominated regime to explore which parameters should be optimized to improve sensitivity. Taking $\Delta(\lambda) \simeq \Delta(\lambda, a_\mrm{min})$ and setting $\lambda = a_\mrm{min}$ to study our sensitivity around where we expect it to be optimal, we find that
\begin{align}
    &\alpha_\mrm{lim} \sim 10^{-4} \left(\frac{a_\mrm{min}}{7\,\mrm{mm}}\right)^2\left(\frac{350\,\mrm{mg}}{m_s}\right)\\
    &\times \left[\left(\frac{40\,\mrm{mg}}{m_p}\right)\left(\frac{1000\,\mrm{h}}{T_\mrm{obs}}\right)\left(\frac{10^7}{Q}\right)\left(\frac{T}{1\,\mrm{K}}\right)\left(\frac{f_s}{35\,\mrm{Hz}}\right)\right]^{1/2}\nonumber\,,
\end{align}
where we have fiducialized to our medium-term configuration. We see that the optimal parameters to consider are the minimal separation distance and the mass of the source.

\section{E.~Choice of Experimental Parameters}

We chose our experimental values to maximize our sensitivity while minimizing the experimental difficulty in their implementation. We justify our proposed parameters below.

The source masses are limited to $350\,\mrm{mg}$ in the medium term to ensure that the size of the (effective) driving mass is well below the minimal $r_0$ and the radius of the dense disk, which limits the effects of the extended nature of the mass on the force gradient.
We also limit the mass of the levitated particle to $400\,\mrm{mg}$. To not change the trapping frequency, only non-magnetic mass is added to the particle between the short- and medium-term stages, as the magnetic mass directly determines the trapping frequency, where the levitation height of the particle is determined by the total mass of the particle and the magnetic mass~\cite{ Timberlake:2021oqf}. This height sets a limit to the amount of non-magnetic mass we can add. In the long-term experiment, to minimize the particle separation and maximize the particle mass, the trap resonance frequency is raised to $f_0 = 150\,\mrm{Hz}$.

Observation times are limited for practical concerns, which in turn also limits the effective $Q$-factor we employ in our detection. Levitated systems have been shown to achieve $Q$-factors of the order of $10^7$ in magnetic systems~\cite{ Timberlake:2021oqf, Fuchs:2023ajk} and in excess of $10^{10}$ using other levitated technologies with lower masses~\cite{PhysRevLett.132.133602}. The proposed temperatures for the short term are easily achievable in a liquid helium bath, which can be further cooled by evaporative cooling to the medium-term proposed temperature of $1\,\mrm{K}$. To reach the long-term temperature, a dilution refrigerator is required.\\

In the short-term setup, our precision is mainly limited by our temperature readout, the frequency stability of the drive, and the precision with which we can determine the sensing mass. Based on a selection of commercially available methods to measure and control these values, the dominant uncertainty is expected to be the temperature. Thermocouples are wildly used for temperature sensing, but suffer from low signal-to-noise ratios at deep cryogenic temperatures, from which we estimate a $10\%$ dominant uncertainty in the short term. Use of commercial resistive and magnetic susceptibility thermometers would reduce this to be subdominant, in the order of $10\,\mrm{\mu K}$ on a proposed temperature of $100\,\mrm{mK}$, for an error of $0.01\%$.

Commercial optical encoders typically have anywhere from $500$ to $100,000$ lines per revolution. On a frequency of $35\,\mrm{Hz}$, our disk spins at roughly $10\,\mrm{Hz}$, or a $100 \mrm{\mu Hz}$ frequency resolution using a $100,000$ line encoder, leading to an error of $0.01\%$, which is subdominant in both short and medium term set-up. In the long term set-up, where a second particle is driven, optical encoders cannot be used, so we conservatively estimate a frequency uncertainty of $0.1\%$, which driven primarily by the phase noise in the system and the performance of typical PLL loops. This is the dominant uncertainty in the long term configuration.

To determine the source masses, commercial mass balance scales provide resolutions of up to $0.1\,\mrm{\mu g}$ on a mass of $10\,\mrm{g}$, resulting in a fractional uncertainty of $10^{-6}\%$. However, for the medium term, our total mass is larger than this total allowed mass. We therefore require less sensitive scales, so we conservatively estimate a dominant uncertainty of $1\%$ in the medium-term configuration arising from the mass uncertainty. 

We note that the mass uncertainty is an important issue in any small-scale gravity measurement, and therefore also for a fifth-force search using these techniques. In a similar fashion to how we turn our positional uncertainty into a systematic error via our pull parameter, which does not play an appreciable role when the slope in our estimator is manifest, we eventually aim to use several different mass wheels with slightly different source masses. When the masses are measured via the same technique (either through balance scales or though SEM volume measures possible at this mass scale), this moves our mass uncertainty to a systematic error. This is also largely eliminated through the use of the slope.

The thermal and frequency stabilities are sources of non-systematic noise that dominate the uncertainty in the pull parameter. Improvements on the thermal stability can be readily made to reduce this noise for the medium- and long-term setups by increasing the resolution of the thermometers used and employing a PID scheme to stabilize the temperature of the experiment. The frequency stability can be increased by using higher resolution optical encoders and dedicated driving electronics for use at low frequencies. At high $Q$-factors, this frequency needs to be stable to better than $f/Q$. For the short- and medium-term setups, this stability can be readily achieved through commercially available optical encoders. For the long-term setup, dedicated electronics must be used to create a stable Phase-Locked Loop.

\clearpage

\end{document}